# Topographical study of TiO$_2$ nanostructure surface for photocatalytic hydrogen production

Gihoon Cha[a], Kiyoung Lee[a], JeongEun Yoo[a], Manuela S. Killian[a], Patrik Schmuki[a,b*]

a. Department of Materials Science and Engineering, WW4-LKO, University of Erlangen-Nuremberg.
b. Department of Chemistry , King Abdulaziz University, Jeddah, Saudi Arabia.

\* To whom correspondence should be addressed.

Email: schmuki@ww.uni-erlangen.de

Tel: +49 9131 852 7575, Fax: +49 9131 852 7582

**Abstract**

In the present work we investigate the photocatalytic hydrogen production (water splitting) activity of different Pt loaded TiO$_2$ nanotube layers. Therefore, we fabricate free standing membranes and fix them in four different configurations: top up (with initiation layer "grass" or without) and bottom up (bottom closed or open), then decorate the tubes with various amounts of Pt and measure the open-circuit photocatalytic H$_2$ production rate. We find a strong influence of the configuration with the open top morphology showing the highest photocatalytic hydrogen production efficiency, these nanotubes yield 3.5 times more H$_2$ than the least efficient structure (bottom closed). The work therefore provides valuable guidelines for optimizing TiO$_2$ nanotube layers for photocatalytic applications.

# 1. Introduction

In search of alternative energy sources the photocatalytic production of hydrogen for energy storage by solar light has gained strongly reignited attention [1-5]. One of the most elegant pathways to produce $H_2$ is the direct splitting of water into $H_2$ and $O_2$ using sunlight and a suitable photocatalyst. Since the pioneering work by Fujishima and Honda [3]on the photoelectrochemical decomposition of water using n-type $TiO_2$ as a photoanode, tremendous research efforts have been directed towards the development of such semiconductor photocatalysts with the aim to achieve a cheap and efficient water splitting process [6-12].

In comparison to $TiO_2$ nanoparticles, anodic $TiO_2$ nanotubes (TiNTs) were shown to possess significantly higher charge collection efficiencies [13]and comparably long electron lifetimes [14]. Additionally, these structures are grown directly from a metallic support, which avoids the need for an immobilization process. Their geometry, i.e., the length, diameter, wall thickness and degree of order can be tuned by variation of the anodization conditions (electrolyte, applied voltage, time) [15,16]. Due to their high versatility, self-organized $TiO_2$ nanotubes are considered for a variety of applications such as solar cells [17,18], photocatalysis [3,19]or biomedical materials [20,21]. The photocatalytic properties of the TiNTs make them particularly interesting for hydrogen production. The highest photocatalytic activity was reported for TiNTs with anatase or rutile crystal structure, which can be obtained by annealing the as grown tubes at temperatures exceeding 300°C and 500°C respectively [15,16].

However, the application of $TiO_2$ structures as photocatalyst for water splitting in most cases [22,23]requires the deposition of a co-catalyst to achieve a reasonable $H_2$ yield. The most significantly enhanced $H_2$ generation rates are obtained by the deposition of Pt nanoparticles on $TiO_2$, as they are able to trap conduction band electrons and to mediate the transfer of the latter to the liquid phase; furthermore, Pt nanoparticles represent efficient catalytic sites for the recombination of atomic hydrogen to $H_2$ [24-28]. In order to achieve a maximum electron transfer yield, additional sacrificial hole capture agents in the electrolyte, such as ethanol, methanol, or glycerol, are used to enhance the overall $H_2$ production rate [15,29].

The decoration of $TiO_2$ structures with noble metal co-catalyst particles commonly is achieved by sol–gel techniques, impregnation, photo- or electro-deposition [30-32] or by anodization of a noble metal − Ti alloy [33,34]. A very homogeneous distribution of the co-catalyst particles can also be obtained by sputter deposition of a thin layer of the noble metal and subsequent conversion to nanoparticles in a thermal dewetting step [35]. The size distribution of the obtained particles is proportional to the thickness of the sputter deposited layer.

Furthermore, the detailed morphology of the TiNTs can easily be expected to be a critical factor in photoelectrochemical applications, e.g., with regard to surface area and light absorption [36,37]. Tubes grown in organic electrolytes possess morphological and composition gradients along the tube length, i.e., they consist of a second, V-shaped inner layer containing significant amounts of carbon [38-40] this layer is most dominant at the tube bottom. The influence of this carbon rich layer on the photocatalytic behavior to date has not been evaluated − mainly because the penetration depth of UV and visible light in the TiNTs typically is too shallow to reach deep enough into the tubes (<1μm for UV light versus typically used tube length of 7-15μm). Recent studies indicate that removal of the carbon rich layer can improve the conductivity of bare TiNTs [41]. Additionally, the influence of the tube top morphology on the photocatalytic behavior, especially residual "grass" on the top surface, is not well investigated either. This "grass" morphology occurs if tubes are grown for extended times in the etching solution and the tube walls at the top (the longest exposed location) get thinned and perforated – the overall morphology then appears as tubes that carry a layer of spikes, i.e. "grass, on top.

Therefore, in the present work, we examine and compare co-catalyst (Pt) aided photocatalytic hydrogen production of different surface morphologies of TiNT layers: tubes i) with and ii) without residual grass on the top surface, and two morphologies where membranes were fabricated and the bottom side of the TiNTs is facing upwards, either with iii) closed or iv) open bottoms. All structures were fabricated as free-standing nanotube layers on FTO, loaded with different amounts of Pt and investigated with regard to their photocatalytic activity in view of the $H_2$ production yields.

**2. Experimental**

For all investigations TiNT layers were produced in an anodic process from Ti sheets (99.6% purity, Advent Materials, UK) of 0.125 mm thickness. Prior to anodization the sheets were cleaned with acetone, ethanol, and distilled water for 15 min each in an ultrasonic bath and then dried in a nitrogen stream.

Electrochemical anodization was carried out at a constant voltage of 60 V for 60min using a DC power supply in two electrode configuration with Pt foil as the counter electrode and the Ti sheets as the working electrode. The electrolyte consisted of 0.15 M $NH_4F$ and 3 vol % $H_2O$ in ethylene glycol (Sigma Aldrich 99.5 %).

In order to form free-standing $TiO_2$ nanotube layers, the TiNTs were detached (see below) from the Ti substrate and attached onto fluorine-doped tin oxide (FTO) glass substrate with doctor-bladed $TiO_2$ nanoparticle paste (Ti-Nanoxide HT, Solaronix SA, Switzerland.

Grass top, open top and bottom closed layers were produced in a 2-step anodization process as illustrated in Figure 1a. After the first anodization, the sample was rinsed in ethanol and then annealed at 350°C for 1 h in a furnace. The partially crystallized sample was anodized in a second step under identical conditions. After the 2nd anodization, the sample was gently shaken and immersed in a weakly acidic solution (0.07 M HF) to separate the two different $TiO_2$ nanotube layers. The upper layer was detached in this process and could be collected. It displays a top morphology with collapsed residues of etched tubes ("grass top"). The grass top TiNTs were annealed for 30 min at 350°C on the FTO substrate and ultrasonicated for 20 min in order to remove the grass and to obtain the "open top" morphology. The lower layer, which was obtained in the 2nd anodization step, was detached by immersing the sample in ethanol, dried and flipped over in order to yield a well-defined bottom closed morphology ("bottom closed").

For the preparation of "bottom open" layers, a so-called potential shocking method was used. The anodization was also carried out at 60V in the same electrolyte (0.15 M $NH_4F$ and 3 vol % $H_2O$ in ethylene glycol) for 60 min and subsequently the anodization potential was rapidly increased to 120V for 3 min. The TiNTs were detached in ethanol, dried and flipped over onto the FTO substrate.

For Pt nanoparticle modification on the layers, Pt was deposited by plasma sputtering (15 mA, $10^{-2}$ mbar, EM SCD500, Leica) with a nominal thickness of 1 nm, 5 nm or 10 nm. All samples were annealed at 500 °C for 1 h in atmosphere in order to transform the Pt layer into Pt nanoparticles by thermally induced dewetting [35]. TiNT samples without Pt particle decoration were annealed under identical conditions for reflectance and $H_2$ production measurements.

The morphological characterization of the different TiNTs was determined by field-emission scanning electron microscopy (Hitachi FE-SEM S4800, Japan) and the crystal structure by X-ray diffraction analysis (XRD, X'pert Philips MPD with a Panalytical X'celerator detector) using graphite-monochromated Cu-Kα radiation ($\lambda = 1.54056$Å). X-ray photoelectron spectroscopy (XPS, PHI 5600, US) using monochromated Al-Kα radiation (1486.6 eV; take-off angle 45°, pass energy 23.5 eV, resolution 0.2 eV) was used to analyze the composition of the samples. The binding energy scale was corrected to the C1s signal at 284.8 eV, the background was subtracted using the Shirley method and atomic compositions were determined with the software MultiPak (V6.1A, 99 June 16, Copyright © Physical Electronics Inc., 1994-1999). To investigate the optical properties of samples, UV-Vis diffuse reflectance spectra (DRS) were measured on a LAMBDA 950 UV-Vis spectrophotometer (Perkin Elmer, Beaconsfield, UK).

Photocatalytic $H_2$ production was evaluated by immersing the samples in a quartz tube containing a 20 vol% methanol solution that was previously purged with with $N_2$ for 20 min to remove $O_2$. Subsequently, the samples were irradiated with UV light for 7h, using a HeCd laser ($\lambda = 325$ nm, 60 mW / cm$^2$, Kimmon, Japan) or a CW-laser ($\lambda = 266$ nm, 11.9 mW / cm$^2$, CryLas GmbH, Germany). In order to measure the $H_2$ production from the methanol solution, gas samples were analyzed by gas chromatography (GCMS-QO2010SE, SHIMADZU).

## 3. Results and discussion

In order to compare different surface morphologies of $TiO_2$ nanotubes, the anodized layers were detached from the Ti substrate and transferred to an FTO substrate that was covered with a layer of $TiO_2$ nanoparticles as illustrated in Fig. 1(a) and described in detail in the experimental section. The $TiO_2$ nanoparticle layer granted improved adhesion, mechanical stability and a good electric contact between the TiNTs and the FTO substrate. Depending on the fabrication approach, four different kinds of surface morphologies were obtained: tubes presenting the top surface with residual collapsed tubes described grass ("grass top") or with the grass removed ("open top"). Also, flipped over tubes showing closed ("bottom closed") and open nanotube bottoms ("open bottom") were investigated. All of the layers have similar dimensions with a diameter of approximately 110 nm and a length of about 15µm (top right insets in Fig.1 (b) − (e)). For the fabrication of free standing $TiO_2$ nanotube layers grown from a Ti metal substrate, most strategies are based on weakening the interface of metal and oxide with chemical or mechanical treatments. However, the as-formed oxide layer also can be damaged by the detachment treatments. To minimize such problems, $TiO_2$ nanotubes were partially crystallized by annealing at 350$^o$C and then anodized again to form a chemically or mechanically weak interface between the two nanotube layers (Fig.1(a, i)). Another strategy to obtain free standing tubes is based on applying a high potential at the end of the anodization − a so-called potential shock, which leads to local acidification or gas evolution at the oxide-metal interface [42,43] and allows to detach the TiNTs with open tube bottoms (Fig.1(a, ii)).

As shown in Fig.1(b) with sufficient anodization time (here: 1h) the nanotube walls are thinned at the top as they are exposed to and attacked by the fluoride containing electrolyte. The thinned tube walls finally collapse and form a "grassy" layer consisting of bundles of nanotube remains. To remove such nanotube-grass the most simple approach is cutting it off by ultrasonication in ethanol, as illustrated in Fig. 1(a, i). The top surface of the ultrasound treated TiNTs is shown in Fig. 1(c). The bottom of this $TiO_2$ nanotube layer is partially damaged due to the second oxide layer grown underneath the first TiNTs and the mechanical stress they experience during detachment (left inset in Fig. 1(c)). It shows a double walled morphology which is common for annealed TiNTs grown in organic electrolytes [15,38]. In contrast, the $TiO_2$ nanotube layer formed in the second anodization step exhibits a well-defined and perfectly closed bottom structure (Fig. 1 (d)).

As illustrated in Fig. 1(a, ii), free standing $TiO_2$ nanotube layers with homogeneously opened tube bottoms (Fig. 1 (e)) can be fabricated by applying a potential shock to the anodized layer.

Compared to the open top TiNTs, the inner diameter is obviously smaller (~ 50 nm) and tubes show smaller spacings, i.e., the outer diameter is slightly larger (~ 120 nm). This finding is in-line with the literature reporting a conical structure (V- shape) for $TiO_2$ nanotubes [15].

In order to use the layers for effective photocatalytic hydrogen production, Pt was sputter deposited onto the surface as a co-catalyst. Fig. 2 shows top and cross-sectional SEM images of the nanotube surfaces decorated with Pt. The Pt is sputtered with a nominal layer thickness of ~5 nm and transforms into nanoparticles homogeneously distributed over the $TiO_2$ surface upon heating. Fig. 2 (a) and (b) show that the TiNTs are homogeneously decorated along the whole surfaces with particles of 2~7 nm size. Furthermore, Pt nanoparticles are deposited into the tubular structures in the case of grass top (inset in Fig.2 (a)) or open top TiNT structures (inset in Fig.2 (b)). The depth distribution of Pt in the grass top TiNTs is a few hundred nm and in the open nanotube structure Pt particles were found up to depths of 2.0 μm by SEM cross-sections.

In contrast, Fig. 2(c) shows that the closed bottom surface is homogeneously covered with a thin Pt layer. No particles were observed in the spacing between the tubes, as can be seen from the inset in Fig. 2(c). In the case of open bottom TiNTs, the Pt nanoparticles are relatively well dispersed (Fig. 2(d)). Due to the closer packing of the tubes with regard to the open top TiNTs and the small inner diameter, Pt particles were hardly observed on the inner and outer walls of the tubes, cf. inset in Fig. 2 (d).

During the annealing process for the dewetting of the Pt particles, the $TiO_2$ nanotube layers were transformed into photocatalytically active anatase. Also, the reference materials without Pt decoration were subject to an identical annealing treatment. Figures 3 (a) and (b) show the crystallinity and crystal orientation of the annealed layers with and without Pt decoration as determined by XRD diffraction. The XRD patterns show virtually the same diffraction peak for all TiNT samples (Fig. 3(a)). Only the grass top and open top structures experienced a mild annealing treatment up to this step in fabrication, which means that the anatase signals are also caused by the TiO$_2$ particles at the interface of tubes and FTO. The anatase peaks at 54° and 55° are more pronounced on the grass top and open top samples, which is in agreement with the additional short annealing treatment that these structures received. After Pt decoration, hardly any shift or peak broadening is observed, additional Pt peaks appear in the, cf. Fig. 3(b). From the results we can assume that Pt particles were successfully loaded onto the TiO$_2$ nanotube layers without any substantial influence on the orientation or lattice spacing of TiO$_2$.

The TiO$_2$ nanotube layers decorated with 5 nm Pt particles additionally were examined by XPS as shown in Fig. 3 (c). The observed amount of Pt is highly depended on layer orientation (bottom up or bottom down). The determined values for the atomic compositions are as follows: grass top: 15.21 at% Pt (Pt/Ti ratio 0.71), open top: 15.43 at% Pt (Pt/Ti ratio 0.73), bottom closed: 30.34 at% Pt (Pt/Ti ratio 2.37) and bottom open: 30.80 at% Pt (Pt/Ti ratio 2.15). The top sides of the tubes show a significantly lower Pt4f signal than the tubes oriented with the bottom upwards. This is in agreement with the SEM investigations, as the Pt particles are distributed along the inside of the tubes in the case of grass top and open top TiNTs, whereas the same amount of Pt is deposited as a layer of agglomerated particles on top of the bottom closed and bottom open tubes, i.e. the relative amount of Pt detected within the information depth of XPS (~3-10 nm) should be significantly higher for the bottom up surfaces. The C2p signal shows a similar trend, which can be ascribed to the V-shaped form of the tubes which have a thicker, carbon-rich inner layer at the bottom of the TiNTs [44].

In order to evaluate some of the optical properties of these structures, diffuse reflectance spectroscopy measurements were conducted on the different TiNTs with and without Pt nanoparticle coatings (Fig. 4(a) – (b)). The bare FTO substrate is included as a reference. From the reference it is clear that the reflectance of the bare FTO substrate drops at around 350 nm as expected due to the band-gap absorption of $SnO_2$. The strong decrease of reflectance in the region of 400 nm can thus be attributed to the increase of light absorption by $TiO_2$ (bandgap of ~3.2 eV). In this region, grass top and bottom closed $TiO_2$ nanotube surface show higher reflectivity than the open top and open bottom surfaces (Fig. 4a). This indicates that the well-ordered open tubular structures influence the light absorption, which is slightly increased for these samples.

While the light absorption of the TiNT surfaces increases in the visible range after Pt nanoparticle decoration (Fig. 4 (b)), in the UV range the diffuse reflectance particularly of the bottom closed and also of the bottom open TiNTs decorated with Pt is obviously increased.

In order to evaluate the photocatalytic activity of $TiO_2$ nanotubes in dependence of the top surface morphology, the photocatalytic $H_2$ production was measured in a solution of 20 vol% of methanol in water upon irradiation for 7 h. Two wavelengths in the UV range (325 nm and 266 nm) were chosen in order to determine if the photocatalytic $H_2$ production is dependent on the used light source. The TiNT samples without Pt decoration all yielded comparable but low efficiencies (not shown). Fig. 4(c) and (d) show the amount of evolved $H_2$ obtained from the different TiNTs with deposited Pt nanoparticles under the two investigated wavelengths, 325 nm and 266 nm, respectively. The results show that the open top morphology produces the highest yield of $H_2$ independent of wavelength and consequently possesses the highest photocatalytic activity. This can be explained by the low reflectance of the material (cf. Fig 4(b)) and the large photocatalytically active surface area caused by the in-depth Pt decoration of the $TiO_2$ nanotube layer (~2μm), which is in-line with the UV light penetration depth [45]. The trend observed for both used wavelengths is comparable, with the 266 nm irradiation producing significantly lower yields, which is in accordance to the power densities of the used lasers.

The grass top and open top morphologies obviously display a higher photocatalytic hydrogen production rate than the bottom up layers (bottom closed and bottom open), which can be ascribed to the available $TiO_2$ surface [35] as well as the optimized depth distribution of Pt on the open top surfaces. Concerning the co-catalyst dispersion, Pt nanoparticles on the bottom up layers are packed much denser than on the bottom down layers, which might hinder the light absorption of $TiO_2$ and decrease the photocatalytic activity. Additionally, the bottom open morphology shows enhanced activity compared to the bottom closed TiNTs, which can also be ascribed to the lower reflectance and the depth distribution and packing density of the Pt nanoparticles.

In order to clarify the influence of the density of the Pt particles, the photocatalytic activity was evaluated in dependence of the Pt particle coverage on the sample with the highest $H_2$ production yield, the open top TiNTs. Two different thicknesses of Pt were sputtered onto the open top TiNTs additionally, i.e., nominal 1 nm and 10 nm, and subsequently heat treated to form a Pt nanoparticle layer. The SEM images show a clear difference in dependence of the sputtered thickness. Only very few Pt particles are loaded by sputtering of a 1 nm thick layer of Pt (Fig 5a), whereas Pt particles cover almost the entire top surface and both the inner and outer shell of the TiNTs in the case of a 10 nm thick sputtered layer of Pt (Fig. 5b).

The amount of $H_2$ produced by irradiation with the 325 nm and 266 nm lasers again shows a comparable trend for the efficiency, which is strongly dependent on the Pt coverage of the TiNT structure. Deposition of only 1 nm Pt leads to a decrease in activity to ~40-45% compared to the 5 nm Pt sample. Also, the deposition of 10 nm leads to a significant decrease in $H_2$ evolution (~70% and 45% respectively), supporting the hypothesis that a too high coverage with Pt shadows the $TiO_2$ and consequently decrease the amount of absorbed UV-light. The results show that the Pt particle density is crucial for the photocatalytic performance of the TiNTs. Optimized photocatalytic activity could be obtained for open top

TiNTs covered by Pt nanoparticles derived from a 5 nm sputter deposited layer of Pt as this geometry facilitates the most favorable distribution of the co-catalyst on the tubes.

## 4. Conclusions

We fabricated four different types of TiNTs, grass top, open top, bottom closed and bottom open, by using either a 2-step anodization or a potential shock method and analyzed the photocatalytic properties according to the top geometry. Open top TiNTs showed improved photocatalytic properties and $H_2$ generation compared to the other investigated structures. This result can be attributed to the beneficial top geometry, in which Pt can be dispersed deeper into the tubular structure. This combination of TiNTs and Pt nanoparticles has an suitable geometry with respect to penetration depth of light. Furthermore, we demonstrated that an optimized coverage with Pt also is an important factor for the photocatalytic performance of TiNTs, Pt nanoparticles formed by dewetting of a 5 nm Pt layer produced the highest $H_2$ yield.

In summary, to maximize the efficiency of the photocatalyst, not only the extent of Pt decoration but also an optimal depth distribution are considered to be very important factors. The distribution of the co-catalyst particles in turn is strongly dependent on the top morphology of the TiNTs, with grass-free, open tubes allowing for the deepest particle decoration and leading to the highest photocatalytic efficiency.

**Acknowledgements**

The authors would like to acknowledge the ERC, the DFG and the DFG cluster of excellence (EAM) for the financial support and H. Hildebrand for the valuable technical help.

**Figure captions**

**Fig. 1** (a) Schematic diagram of the fabrication process of i) grass top, open top, bottom closed, and ii) bottom open TiNTs. Scanning electron microscopy (SEM) images of (b) grass top, (c) open top (inset in bottom left corner: bottom view of open top TiNTs), (d) bottom closed, and (e) bottom open TiNTs. The insets in the top right corner show the cross sections of the tubes.

**Fig. 2** SEM top view images of (a) grass top, (b) open top, (c) bottom closed, and (d) bottom open TiNTs with sputter-deposited Pt particles. The insets show cross-sectional views of the upper part of the structures.

**Fig. 3** (a) X-ray diffraction patterns of grass top, open top, bottom closed, and bottom open TiNTs (annealed) and (b) X-ray diffraction patterns after Pt decoration. The annotations refer to: S – Sn, A – anatase, Pt – platinum. (c) XPS spectra of Pt4f area for Pt decorated grass top, open top, bottom closed, and bottom open TiNTs; (d) atomic concentrations as determined by XPS.

**Fig. 4** (a) Reflectance measurements of grass top, open top, bottom closed, and bottom open TiNTs before Pt sputtering and (b) after (all samples were annealed). The insets show reflectivity for a wider range from 250 to 800 nm wavelength. (c) and (d): Photocatalytic $H_2$ production in a 20 vol.% methanol aqueous solution for Pt (5 nm) decorated grass top, open top, bottom closed, bottom open TiNTs: illumination by (c) 325 nm laser and (d) 266 nm laser.

**Fig. 5** SEM images of open top TiNTs decorated with (a) 1 nm Pt and (b) 10 nm Pt. The insets show cross-sectional views of the upper part of the tubes. (c) and (d): Photocatalytic $H_2$ production in a 20 vol.% methanol aqueous solution for open top TiNTs with different nominal thickness of Pt; illumination by (c) 325 nm laser and (d) 266 nm laser.

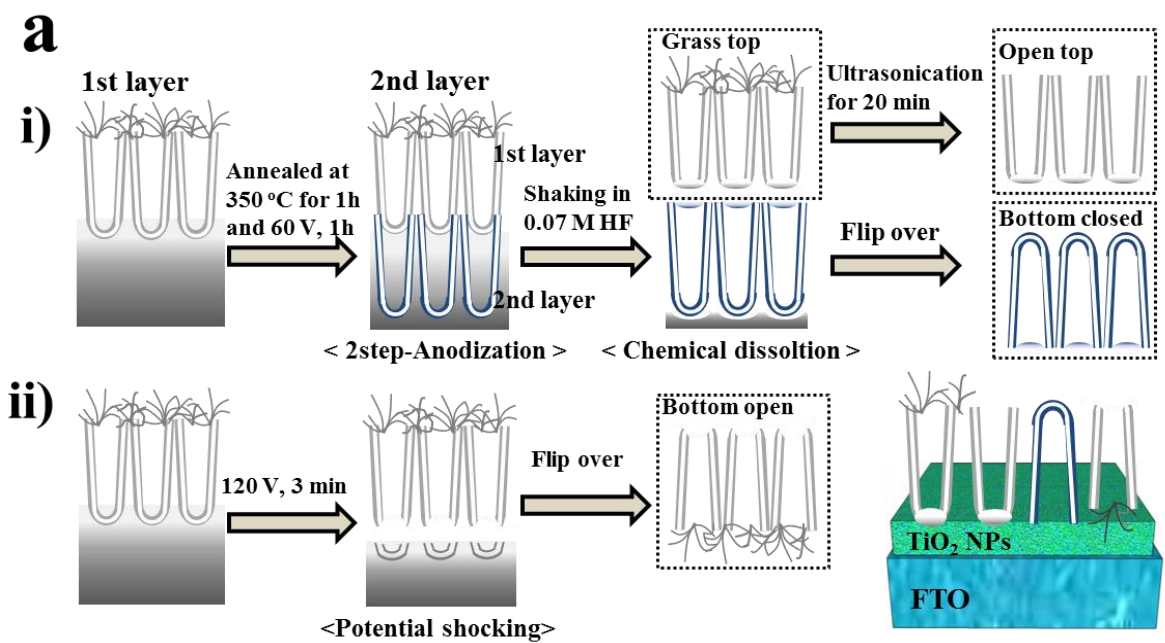
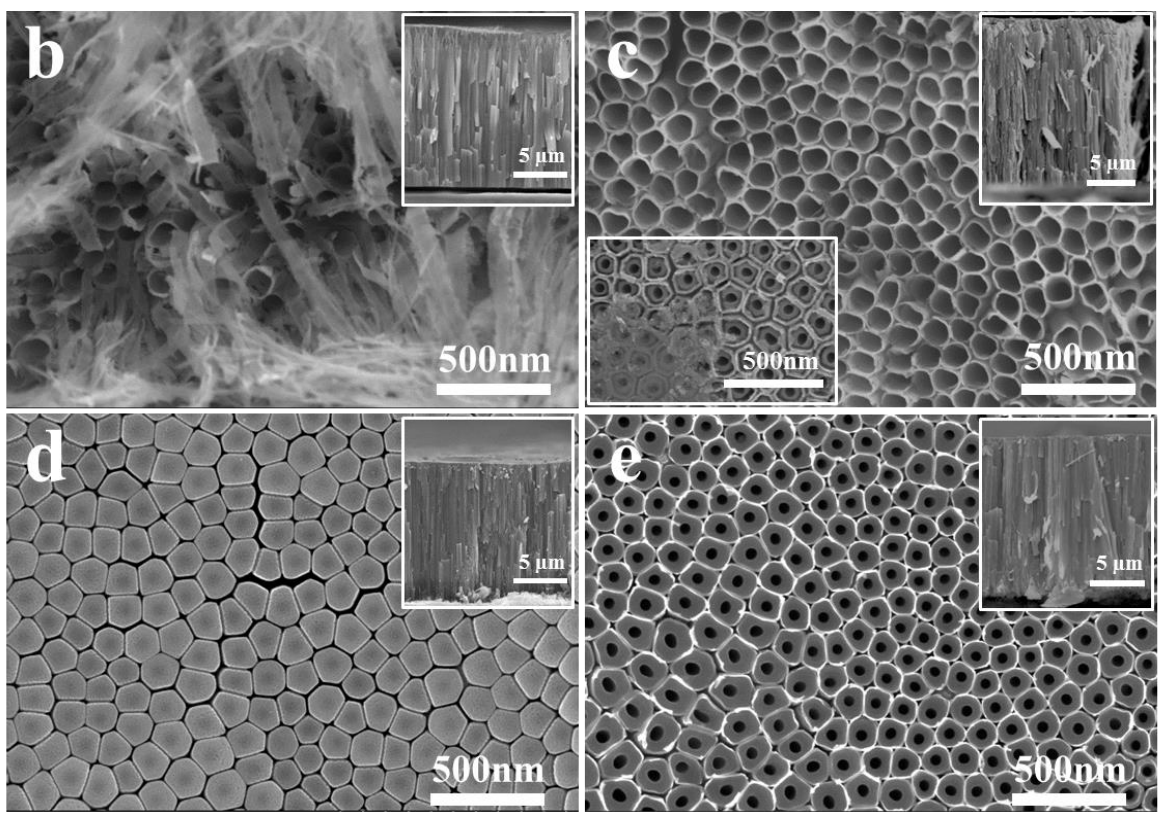

Fig. 1

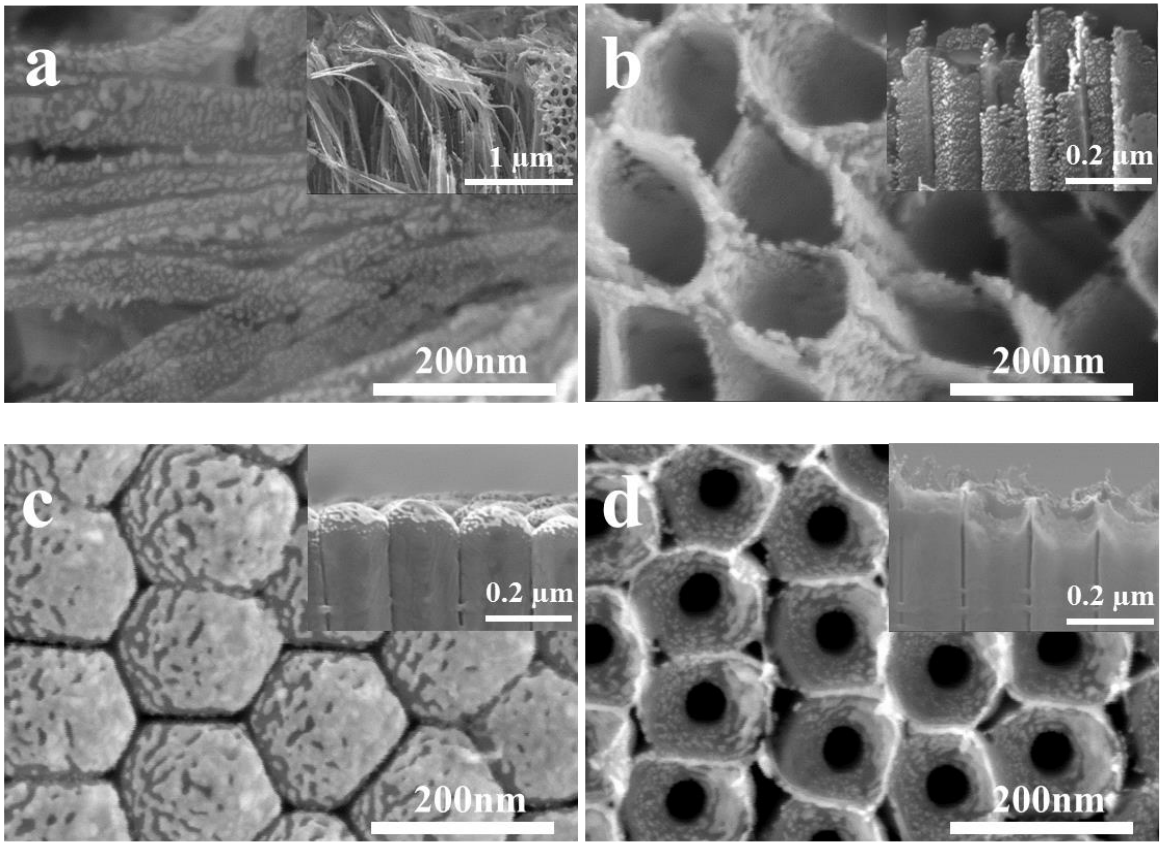

Fig. 2

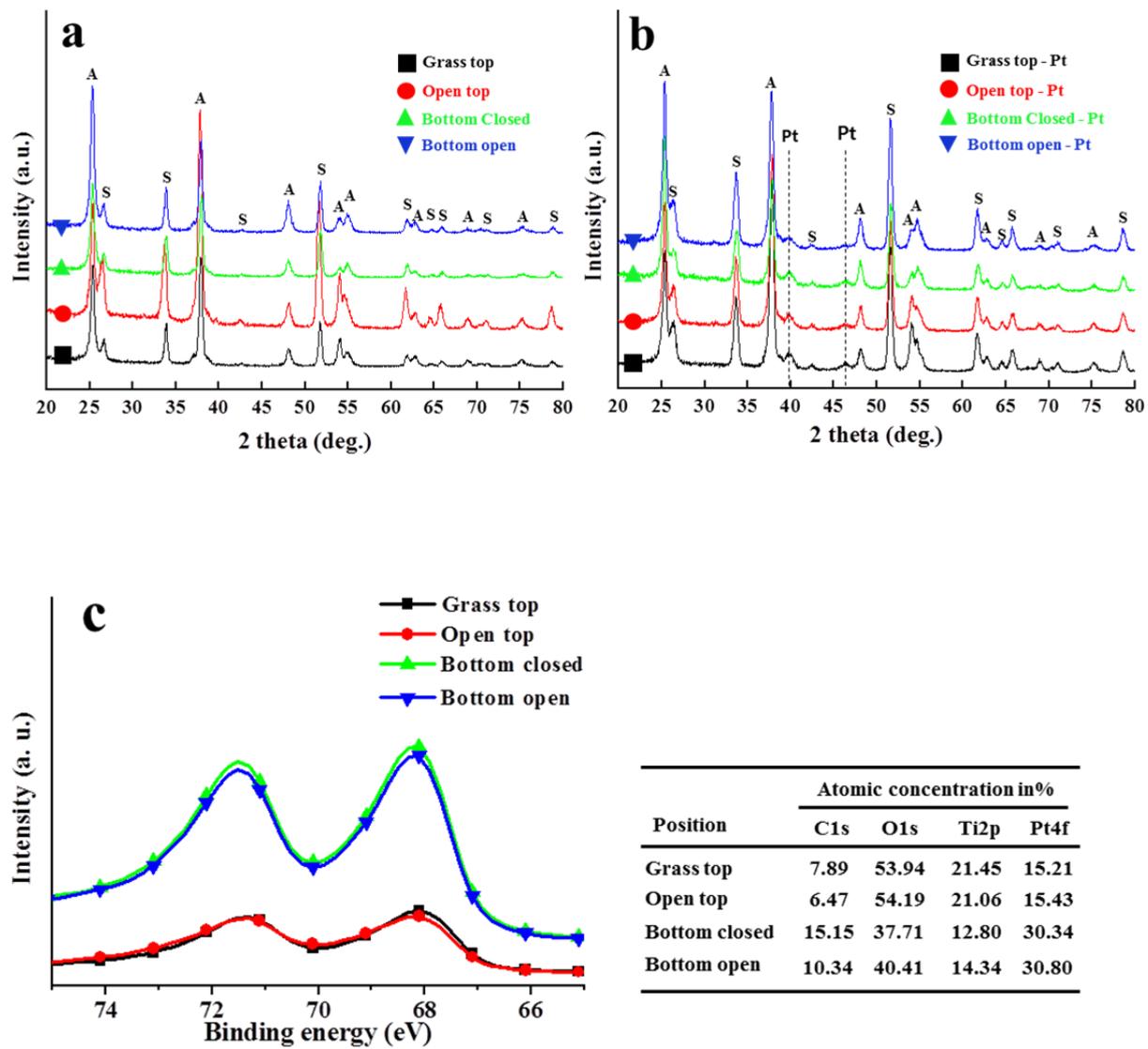

Fig. 3

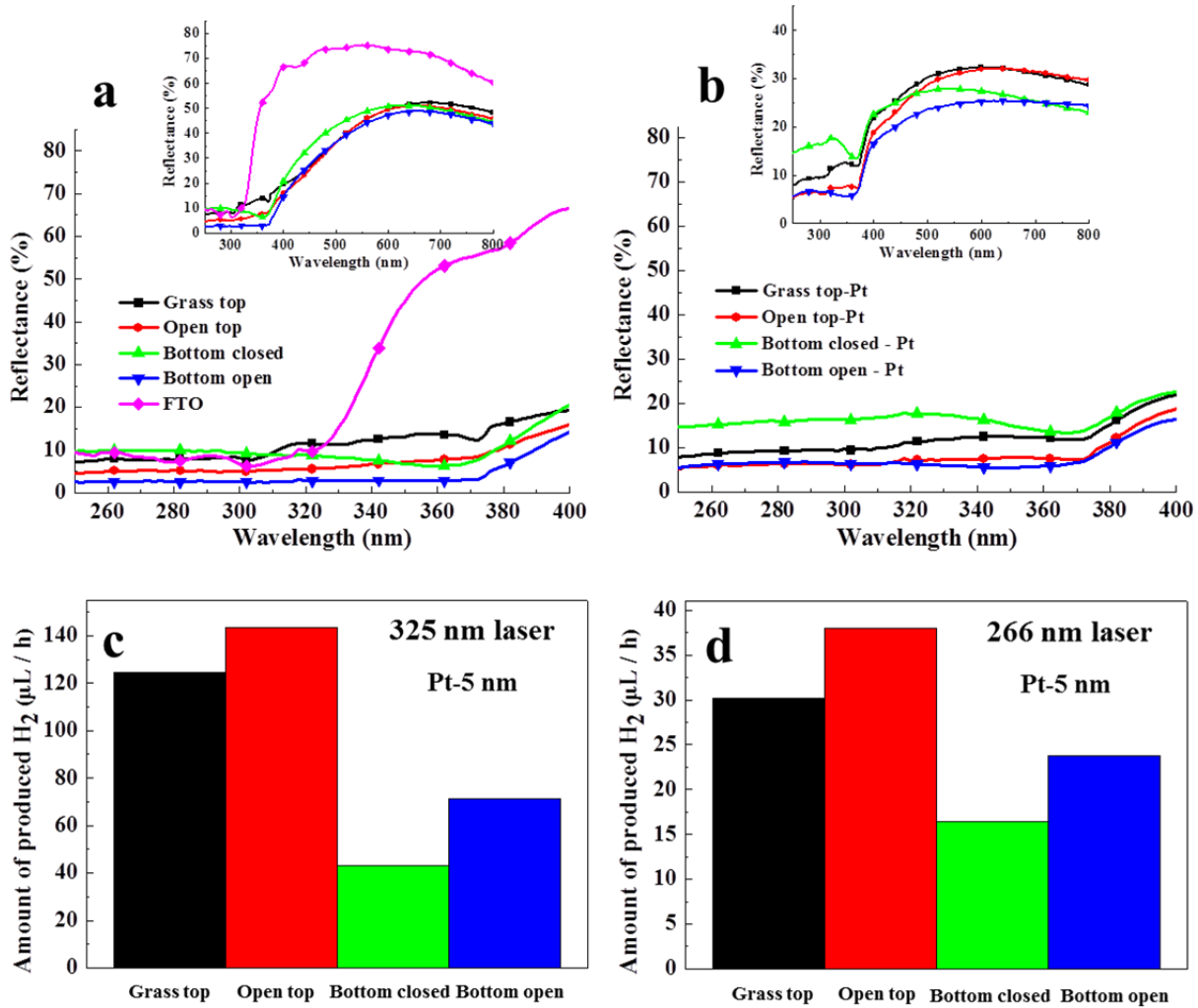

Fig. 4

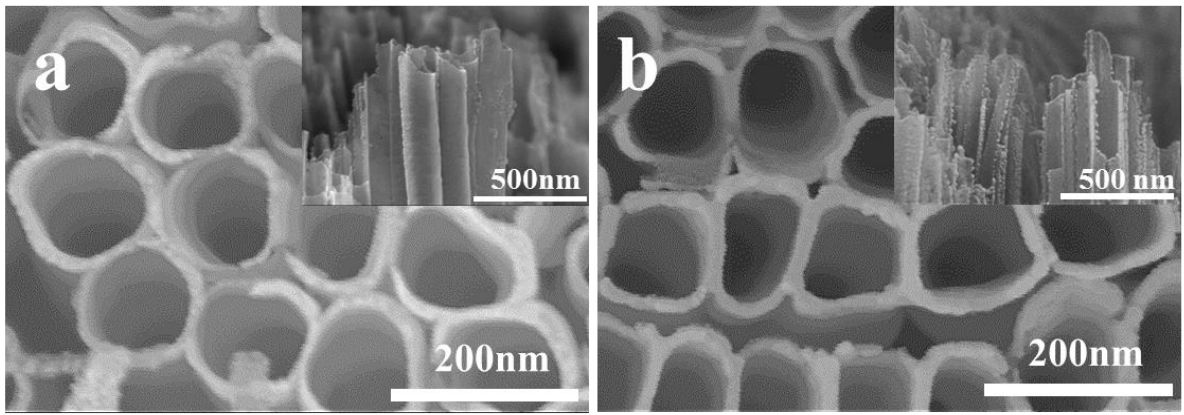

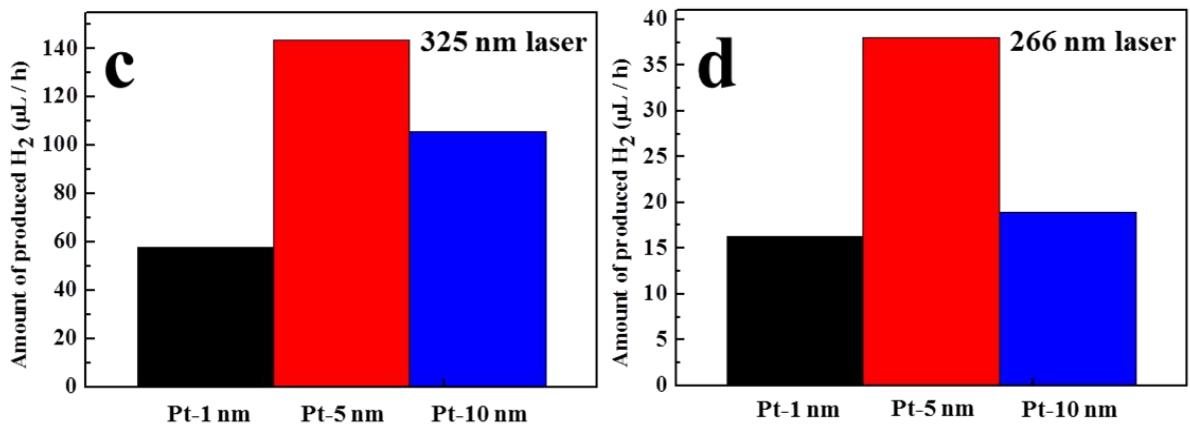

Fig. 5